\documentclass{PoS}
\usepackage{graphicx,amsmath,amssymb,bm}
\usepackage{epsfig}
\title{Three Fermions in a Box}

\ShortTitle{Three Fermions in a Box}

\author{\speaker{Thomas Luu}\thanks{In collaboration with W. Detmold
    and A. Walker-Loud. LLNL-PROC-407504}\\ Lawrence
  Livermore National Laboratory\\ E-mail: \email{tluu@llnl.gov}}

\abstract{I calculate finite-volume effects for three identical
  spin-1/2 fermions in a box assuming short-ranged repulsive interactions of
  `natural size'.  This analysis employs standard perturbation theory
  in powers of 1/L, where L$^3$ is the volume of the box.  I give
  results for the ground states in the $A_1$, $T_1$, and $E$ cubic
  representations. }

\FullConference{The XXVI International Symposium on Lattice Field Theory\\
		 July 14-19 2008\\
		 Williamsburg, Virginia, USA}

\begin{document}

\section{Introduction\label{sect:intro}}
Recently much progress has been made in calculating two-body
low-energy constants (LECs) directly from QCD in the mesonic sector
(see \cite{Beane:2008dv} for a recent review).  Lattice QCD (LQCD)
calculations of two-meson interacting energies are performed and,
using L$\ddot{\mbox{u}}$scher's formalism\cite{Luscher:1986}, the
scattering lengths are extracted.  The exceptionally clean signals
obtained in the mesonic sector have allowed for the extraction of pure
three-body LECs \cite{Beane:2007es,Detmold:2008yn} using the
multi-boson finite volume effects derived in \cite{Beane:2007qr,
  Detmold:2008gh}.  In the three-pion sector, the extracted three-body
LEC is consistent with a repulsive three-body
interaction\cite{Beane:2007es}.

L$\ddot{\mbox{u}}$scher's formalism can be used to extract scattering
phase shifts of two interacting baryons just as in the mesonic
sector\cite{Beane:2006mx}.  However, due to Pauli's exclusion
principle, there is no general formula for relating LECs to
interacting energy shifts for three (and more) fermions.  In this
proceeding I present finite volume effects for three identical
spin-1/2 particles within a box, thereby generalizing
L$\ddot{\mbox{u}}$scher's formalism to three fermions.  This analysis
uses repulsive, short-ranged interactions of `natural size', which is
amenable to perturbation theory.  Results are given for T$_1$ states
that are accurate to order 1/L$^5$, whereas results for A$_1$ and E
states are accurate to order 1/L$^4$.  Here L is the length of a side
of the box.

In the next section I give a heuristic explanation of how to
construct three-fermion states of definite cubic symmetry.  The
anti-symmetry restrictions makes this construction non-trivial.
Section~\ref{sect:perturbation} then enumerates the perturbative
results in powers of 1/L using these basis states.  I conclude in
sect.~\ref{sect:conclusion}.

\section{Constructing anti-symmetrised three-fermion states of good
  cubic symmetry\label{sect:states}}
  
\subsection{Jacobi basis\label{sect:jacobi basis}}
The three-body single-particle eigenstates of the dimensionless
kinetic energy operator $\hat{T}$ (in units of
$\epsilon_0=4\pi^2/mL^2$) in a box of volume L$^3$ are given by
$|\vec{n}_1\ \vec{n}_2\ \vec{n}_3>$, where
\begin{equation}\label{eqn:single-particle eigenstates}
\hat{T}|\vec{n}_1\ \vec{n}_2\ \vec{n}_3>=|\vec{n}_1\ \vec{n}_2\
\vec{n}_3>
\left(\frac{\vec{n}_1^2}{2}+\frac{\vec{n}_2^2}{2}+\frac{\vec{n}_3^2}{2}\right)\ .
\end{equation}
Here $\vec{n}_i=(n_{ix},n_{iy},n_{iz})$ represents the wave number
vector for the $i^{th}$ particle and I have assumed all particles
have equal mass $m$.  Other quantum numbers, such as spin (and
isospin), have been suppressed.

\begin{figure}
\centering
\includegraphics[width=.8\textwidth]{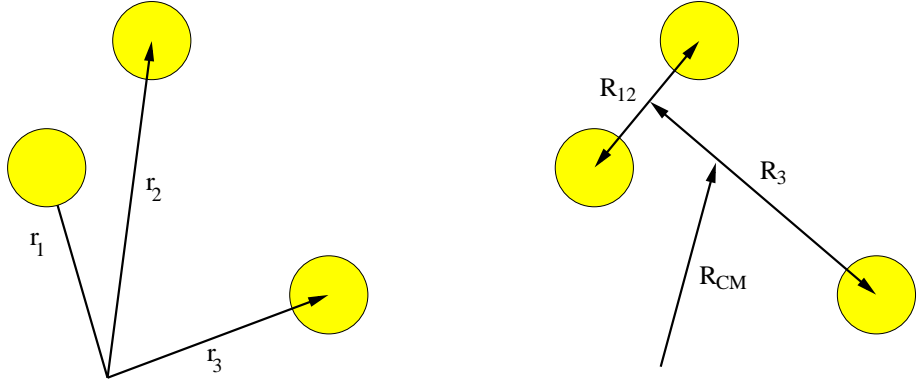}
\caption{Depiction of transformation from independent degrees of
  freedom in the single-particle basis (left) to the independent
  degrees of freedom in the Jacobi basis (right).\label{fig:jacobi}}
\end{figure}

For reasons which will become apparent below, the single-particle
states are now transformed to a Jacobi basis using
\begin{eqnarray}
\vec{R}_{12}&=&\vec{r}_1-\vec{r}_2\nonumber\\
\vec{R}_3&=&\vec{r}_3-\frac{1}{2}\left(\vec{r}_1+\vec{r}_2\right)\\ \label{eqn:jacobi
  r}
\vec{R}_{cm}&=&\frac{1}{3}\left(\vec{r}_1+\vec{r}_2+\vec{r}_3\right)\
.\nonumber
\end{eqnarray}
Here $\vec{R}_{12}$ represents the relative motion between particles 1
and 2, $\vec{R}_3$ represents the relative motion between particle 3
and the center-of-mass (CM) of particles 1 and 2, and $\vec{R}_{cm}$
is the total CM motion.  In momentum space, this corresponds to the
following transformations,
\begin{eqnarray}
\vec{P}_{12}&=&\frac{1}{2}\left(\vec{p}_1-\vec{p}_2\right)\nonumber\\ \vec{P}_3&=&\frac{2}{3}\left(\vec{p}_3-\frac{1}{2}\left(\vec{p}_1+\vec{p}_2\right)\right)\\ \label{eqn:jacobi
  p} \vec{P}_{cm}&=&\vec{p}_1+\vec{p}_2+\vec{p}_3\ .\nonumber
\end{eqnarray}
Eigenstates in this Jacobi basis are defined as $|\vec{N}_{12}\
\vec{N}_3\ \vec{N}_{cm}>$, where now
\begin{equation}\label{eqn:jacobi-basis eigenstates}
\hat{T}|\vec{N}_{12}\ \vec{N}_3\ \vec{N}_{cm}>
=|\vec{N}_{12}\ \vec{N}_3\ \vec{N}_{cm}>
\left(\vec{N}_{12}^2+\frac{3}{4}\vec{N}_3^2+\frac{1}{6}\vec{N}_{cm}^2\right)\ .
\end{equation}
Figure~\ref{fig:jacobi} shows this transformation schematically.  

Since interactions occur only between particles (I assume no external
potential acting on the fermions) and typically only the lowest energy
eigenstates are of interest, the utility of switching to the Jacobi
basis is now manifest: one can simply set $\vec{N}_{cm}=0$.  Thus a
three-body problem effectively becomes a `two-body' problem.  All
results in the following section use $\vec{N}_{cm}=0$, though it is
straightforward (but tedious) to generalize to nonzero CM motion.

A subtle point comes about from the transformation to Jacobi basis
since the box boundary conditions are originally defined in the
single-particle basis.  In the case when $\vec{N}_{cm}=0$, for any
component of $\vec{N}_3$ which is odd, the corresponding component of
$\vec{N}_{12}$ must satisfy anti-periodic boundary conditions.
Conversely, any component of $\vec{N}_3$ which is even has the
corresponding component of $\vec{N}_{12}$ satisfying periodic boundary
conditions.  These restrictions can be derived by comparing the
completeness relations within the single-particle basis and the Jacobi
basis.

\subsection{Three-body antisymmetrised states\label{sect:antisymmetrization}}
The states $|\vec{N}_{12}\ \vec{N}_3>$ (the index $\vec{N}_{CM}$ is
dropped since only zero CM motion is considered) are not
anti-symmetric under exchange of any two particles.  Anti-symmetric
states are constructed by projecting onto the three-body
anti-symmetriser,
\begin{equation}\label{eqn:antisymmetriser}
P^{123}_{\mathcal A}=\frac{1}{3}P^{12}_{\mathcal A}\left(1-P_{13}-P_{23}\right)\ ,
\end{equation}
where $P_{ij}$ is the permutation operator that permutes particles $i$
and $j$ and
\begin{equation}\label{eqn:2bodyantisymmetriser}
P^{12}_{\mathcal A}=\frac{1}{2}\left(1-P_{12}\right) 
\end{equation}
is the two-body anti-symmetriser of particles 1 and 2.  Note that
Eq.~\ref{eqn:antisymmetriser} commutes with the parity operator.  Thus
the states of interest are ones that satisfy
\begin{displaymath}
P_{\mathcal A}^{123}|n\ i\ \pi>=|n\ i\ \pi>\ ,
\end{displaymath}
where $|n\ i\ \pi>$ represents the $i^{th}$ state of cubic shell $n$
with definite parity $\pi$.  It is made up of linear combinations of
$|\vec{N}_{12}\ \vec{N}_3>$ such that $\vec{N}_{12}^2+\vec{N}_3^2=n$.
For each cubic shell there is a finite number of anti-symmetric
states $D_n$.
  
\subsection{Cubic group symmetry\label{sect:cubic group}}
The states $|n\ i\ \pi>$ are now anti-symmetrised but are not states
of good cubic symmetry.  Standard group-theoretical techniques can be
used to obtain the appropriate linear combinations of $|n\ i\ \pi>$
such that the overall anti-symmetric state falls into one of the five
irreducible representations (irreps) of the cubic group: A$_1$, A$_2$,
E, T$_1$, and T$_2$\cite{Slater:1963}.  A cursory description of this
procedure is only given below.

Given the set of anti-symmetrised states $|n\ i\ \pi>$, matrix
elements of the cubic rotation operators $R_\alpha$ are constructed
using this basis, \emph{i.e.}  $<n\ j\ \pi|R_\alpha|n\ i\ \pi>$,
forming a regular representation of the group.  This regular
representation consists of 24 matrices, all of dimension $D_n\times
D_n$.  Traces of these matrices give five distinct characters
$\chi_R(r)$, and using the five characters of the irreps of the cubic
group, $\chi_{IR}(r)$, and the number of rotation elements in each
irrep, $n_{IR}(r)$, the multiplicity of each irrep in this regular
representation,
\begin{displaymath}
m_{IR}=\frac{1}{24}\sum_{r=1}^5n_{IR}(r)\chi_R(r)\chi_{IR}(r)\\ ,
\end{displaymath}
for a given cubic shell $n$ is found.  Given the dimension of each
irrep, $d_{IR}(r)$, projection operators for each irrep are then
constructed,
\begin{displaymath}
P_{IR}=\frac{1}{24}\sum_rd_{IR}(r)\chi_{IR}(r)\sum_{\alpha\in r}R_\alpha\ ,
\end{displaymath}
from which the anti-symmetrised states of definite cubic symmetry are
constructed.  Table~\ref{tab:dimensions} enumerates the anti-symmetric
states for the first three cubic shells.  Note that Pauli's exclusion
principle prevents any three spin-1/2 fermion states residing in the
$n=0$ cubic shell.
\begin{table}
\centering
\begin{tabular}{c||c|c|c|c|c|c|c||c}
\hline 
\hline
$n$ & \quad Spin\ \ \ \  & \quad Parity\ \ \ \
& \quad A$_1$\  \ \ \  & \quad A$_2$\ \ \ \  & \quad E\ \ \ \  & \quad
T$_1$\ \ \ \  & \quad T$_2$\ \ \ \ 
& $D_n$\\
\hline
1 & $\frac{1}{2}$ & + & 1 & 0 & 1 & 0 & 0 & 3\\
\hline
1 & $\frac{1}{2}$ & - & 0 & 0 & 0 & 1 & 0 & 3\\
\hline
1 & $\frac{3}{2}$ & - & 0 & 0 & 0 & 1 & 0 & 3\\
\hline
2 & $\frac{1}{2}$ & + & 2 & 1 & 3 & 1 & 2 & 18\\
\hline
2 & $\frac{1}{2}$ & - & 0 & 0 & 0 & 3 & 3 & 18\\
\hline
2 & $\frac{3}{2}$ & + & 0 & 1 & 1 & 1 & 0 & 6\\
\hline
2 & $\frac{3}{2}$ & - & 0 & 0 & 0 & 2 & 2 & 12\\
\hline
3 & $\frac{1}{2}$ & + & 4 & 0 & 4 & 3 & 6 & 39\\
\hline
3 & $\frac{1}{2}$ & - & 0 & 3 & 3 & 7 & 3 & 39\\
\hline
3 & $\frac{3}{2}$ & + & 1 & 1 & 1 & 2 & 2 & 16\\
\hline
3 & $\frac{3}{2}$ & - & 0 & 3 & 1 & 4 & 1 & 20\\
\hline
\hline
\end{tabular}
  \caption{Dimension and multiplicity of anti-symmetric states of
    various cubic irreps for zero CM motion.  $n$ refers to value of
    cubic shell.  The cubic irreps are A$_1$, A$_2$, E, T$_1$, and
    T$_2$ and refer to the spatial part of the wavefunctions.  Numbers
    below these irreps correspond to the mulitplicity of the irrep
    within cubic shell $n$.  Last column gives the total dimension of
    anti-symmetric states $D_n$ in cubic shell
    $n$. \label{tab:dimensions}}
\end{table}
  
\section{Perturbative results\label{sect:perturbation}}
At up to order 1/L$^5$, only s-wave and p-wave interactions contribute.  I
parametrize these momentum space interactions in the following manner:
\begin{eqnarray}
V_0(\vec{p}',\vec{p})&=&\frac{4\pi
  a_0}{m}\left[1+\frac{a_0r_0}{2}\left(\frac{p'^2+p^2}{2}\right)+.\ .\ .\right]
\quad\ \quad\quad\mbox{(s-wave)}\label{eqn:swave}\\ V_1(\vec{p}',\vec{p})&=&\frac{12\pi
  a_1}{m}\vec{p}'\cdot\vec{p}
\left[1+\frac{a_1r_1}{2}\left(\frac{p'^2+p^2}{2}\right)+.\ .\ .\right]\quad\mbox{(p-wave)}\ .\label{eqn:pwave}
\end{eqnarray}
The parameters $a_0$ and $r_0$ are the scattering length and effective
range, respectively.  They both have units of length.  The parameters
$a_1$ and $r_1$ are the scattering volume and effective momentum,
having units of length$^3$ and length$^{-1}$, respectively.  The
perturbative analysis assumes that $a_0/L\ll 1$ and $r_0/L\ll 1$, as
well as $a_1/L^3\ll1$ and $r_1L\ll1$.

The results, when expressed with dimensional units, are accurate to
order 1/L$^5$ for the T$_1$ states.  For E and A$_1$ states, the
results are accurate to order 1/L$^4$. However, since results are
presented in units of $\epsilon_0=\frac{4\pi^2}{mL^2}$ (\emph{i.e.}
results are dimensionless), at most terms of order 1/L$^3$ are shown
explicitly.  Only states perturbatively connected to the first cubic
shell $n=1$ are shown.  To facilitate the presentation, a list of the
various lattice sums and their numerical values that are inherent to
these calculations is given in tab.~\ref{tab:sums}.

\subsection{T$_1^-$ Spin=$\frac{3}{2}$}
This channel is only sensitive to the effective range,
\begin{equation}\label{eqn:T1m S3/2}
\frac{\epsilon}{\epsilon_0}=1+36\pi\frac{a_1}{L^3}+{\mathcal O}(L^{-5})\ .
\end{equation}
Furthermore, there are no terms that come in at 1/L$^4$ on the
right-hand side of eq.~\ref{eqn:T1m S3/2}.

\subsection{T$_1^-$ Spin=$\frac{1}{2}$}
\begin{multline}\label{eqn:T1m S1/2}
\frac{\epsilon}{\epsilon_0}=1+3\frac{a_0}{\pi
  L}+\left(\frac{3}{2}-3{\mathcal L}_2\right)\frac{a_0^2}{\pi^2
  L^2}+27\pi \frac{a_1}{L^3}+\frac{3}{2}\pi \frac{a_0^2
  r_0}{L^3}\\ +\left(\frac{9}{4}-3{\mathcal L}_2+3{\mathcal
  L}_2^2-6{\mathcal M}_2\right)\frac{a_0^3}{\pi^3 L^3}
+{\mathcal O}(L^{-4})\ .
\end{multline}

\subsection{E$^+$ Spin=$\frac{1}{2}$}
\begin{multline}\label{eqn:Ep S1/2}
\frac{\epsilon}{\epsilon_0}=1+\frac{a_0}{\pi
  L}-\left(\frac{3}{2}+{\mathcal L}_2\right)\frac{a_0^2}{\pi^2
  L^2}+{\mathcal
  O}(L^{-3})\ .\\
\end{multline}

\subsection{A$_1^+$ Spin=$\frac{1}{2}$}
\begin{multline}\label{eqn:A1p S1/2}
\frac{\epsilon}{\epsilon_0}=1+7\frac{a_0}{\pi
  L}-\left(\frac{3}{2}+{\mathcal L}_2+6{\mathcal
  L}_1\right)\frac{a_0^2}{\pi^2 L^2} +{\mathcal O}(L^{-3})\ .\\
\end{multline}

\begin{table}
\centering
\begin{tabular}{c|c|c}
\hline 
\hline
Label & Expression & Numerical value\\
\hline
${\mathcal L}_1$ & $\sum^{|\vec{n}|\le\Lambda} \frac{1}{\vec{n}^2-1}-4\pi\Lambda$ & -1.21134 \\
\hline
${\mathcal L}_2$ & $\sum^{|\vec{n}|\le\Lambda} \frac{1}{n_x^2+n_y^2+n_z(n_z+1)}-4\pi\Lambda$ & -6.37481 \\
\hline
${\mathcal M}_1$ & $\sum \frac{1}{(\vec{n}^2-1)^2}$ & 23.24322 \\
\hline
${\mathcal M}_2$ & $\sum \frac{1}{(n_x^2+n_y^2+n_z(n_z+1))^2}$ & 18.3 \\
\hline
\hline
\end{tabular}
 \caption{Lattice sums and their numerical values.  Sums are over all
   triplet of integers $\vec{n}=(n_x, n_y, n_z)$ such that the
   denominator does not vanish.  The limit $\Lambda\rightarrow\infty$
   is implicit.\label{tab:sums}}
\end{table}
\section{Conclusion\label{sect:conclusion}}
I have presented finite volume effects for three identical spin-1/2
fermions in a box interacting via short-ranged, repulsive interactions
of `natural' size.  Results are given for states in the first cubic
shell $n=1$ and are valid up to order 1/L$^5$ for the T$_1$ states and
1/L$^4$ for the A$_1$ and E states.  These results generalize
L$\ddot{\mbox{u}}$scher's formalism to three spin-1/2 fermions in a
box.

A similar analysis can be performed on nucleons by the introduction of
isospin degrees of freedom\cite{Luu:2008}.  Here the spectra of states
is extremely rich and the structure of the interactions is complex due
to the presence of tensor forces and pure s-wave three-body
interactions.  Furthermore, at the physical pion mass the interactions
are no longer of `natural' size and non-perturbative formalisms must
be employed\cite{Luu:2008}.

Ultimately, LQCD will answer current outstanding nuclear physics
questions, such as the nature and origin of the tensor force and
three-nucleon interaction.  This work represents a necessary step
towards obtaining these answers.

\begin{acknowledgments}
This work was performed under the auspices of the U.S. Department of
Energy by Lawrence Livermore National Laboratory under Contract
DE-AC52-07NA27344.  I thank WD and AWL for insightful discussions on
this work.
\end{acknowledgments}


\end{document}